\documentclass[epj]{webofc}
\usepackage[varg]{txfonts}   
\usepackage{graphicx,color} 
\usepackage{bm}       
\usepackage{amsmath}
\usepackage{amssymb}
\woctitle{21st International Conference on Few-Body Problems in Physics}
\begin{document}
\title{Strong interaction studies with kaonic atoms}
%
%


\abstract{
The strong interaction of antikaons (K$^{-}$) with nucleons and nuclei in the low-energy regime represents
an active research field connected intrinsically with few-body physics. There are important open questions like the question of antikaon nuclear bound states - the prototype system being K$^{-}$pp.
A unique and rather direct experimental access to the antikaon-nucleon scattering lengths is provided by precision X-ray spectroscopy of transitions in low-lying states of light kaonic atoms like kaonic hydrogen isotopes. In the SIDDHARTA experiment at the electron-positron collider DA$\Phi$NE of LNF-INFN we measured the most precise values of the strong interaction observables, i.e.  the strong interaction on the 1s ground state of the electromagnetically bound K$^{-}$p atom leading to a hadronic shift $\epsilon_{1s}$ and
a hadronic broadening $\Gamma_{1s}$ of the 1s  state.
The SIDDHARTA result triggered new theoretical work which achieved major progress in the understanding of the low-energy strong interaction with strangeness. Antikaon-nucleon scattering lengths have been calculated constrained by the SIDDHARTA data on kaonic hydrogen.
For the extraction of the isospin-dependent scattering lengths a measurement of the hadronic shift and width of kaonic deuterium is necessary. Therefore,
new X-ray studies with the focus on kaonic deuterium are in preparation (SIDDHARTA2). Many improvements in the experimental setup will allow to measure kaonic deuterium which is challenging due to the anticipated low X-ray yield. Especially important are the data on the X-ray yields of kaonic deuterium extracted from a exploratory experiment within SIDDHARTA.

}

\author{J.~Marton\inst{1}\fnsep\thanks{\email{johann.marton@oeaw.ac.at}} \and
M.~Bazzi\inst{2} \and
G.~Beer\inst{3} \and
C.~Berucci\inst{1,2} \and
D.~Bosnar\inst{11} \and
A.M.~Bragadireanu\inst{1,5} \and
M.~Cargnelli\inst{1} \and
A.~Clozza\inst{2} \and
C.~Curceanu \inst{2} \and
A.~d'Uffizi\inst{2} \and
C.~Fiorini\inst{4} \and
F.~Ghio\inst{6} \and
C.~Guaraldo\inst{2} \and
R.~Hayano\inst{7} \and
M.~Iliescu\inst{2} \and
T.~Ishiwatari\inst{1} \and
M.~Iwasaki\inst{8} \and
P.~Levi Sandri\inst{2} \and
S.~Okada\inst{8} \and
D.~Pietreanu\inst{2,5} \and
K.~Piscicchia\inst{2} \and
T.~Ponta\inst{5} \and
R.~Quaglia\inst{4} \and
A.~Romero Vidal\inst{10} \and
E.~Sbardella\inst{2} \and
A.~Scordo\inst{2} \and
H.~Shi\inst{2} \and
D.L.~Sirghi\inst{2,5} \and
F.~Sirghi\inst{2,5} \and
H.~Tatsuno\inst{12} \and
O.~Vazquez Doce\inst{9} \and
E.~Widmann\inst{1} \and
J.~Zmeskal\inst{1}
}

\institute{
Stefan-Meyer-Institut f\"{u}r subatomare Physik, Boltzmanngasse 3, 1090 Wien, Austria
\and
INFN, Laboratori Nazionali di Frascati, C.P. 13, Via E. Fermi 40, I-00044 Frascati (Roma), Italy
\and
Dep. of Physics and Astronomy, University of Victoria, P.O.Box 3055, Victoria B.C. Canada V8W3P6
\and
Politecnico di Milano, Dip. di Elettronica e Informazione, Piazza L. da Vinci 32, I-20133 Milano, Italy
\and
Institutul National pentru Fizica si Inginerie Nucleara Horia Hulubei, Reactorului 30, Magurele, Romania
\and
INFN Sez. di Roma I and Instituto Superiore di Sanita I-00161, Roma, Italy
\and
University of Tokyo,7-3-1, Hongo, Bunkyo-ku,Tokyo, Japan
\and
RIKEN, Institute of Physical and Chemical Research, 2-1 Hirosawa, Wako, Saitama 351-0198 Japan
\and
Excellence Cluster Universe, Tech. Univ. M\"{u}nchen, Boltzmannstra{\ss}e 2, D-85748 Garching, Germany
\and
Universidade de Santiago de Compostela, Casas Reais 8, 15782 Santiago de Compostela, Spain
\and
Department of Physics, Faculty of Science, University of Zagreb, Croatia
\and
High Energy Accelerator Research Organization (KEK), Tsukuba, 305-0801, Japan
}

\maketitle
\section{Introduction}
\label{intro}

Hadronic atoms like pionic and kaonic atoms are extremely valuable systems for the investigation of the strong interaction in the low-energy domain.

Especially interesting is the strong interaction involving the strange quark which plays a peculiar role.
The strange quark belongs to the light quarks but it is with a mass of about 100 MeV/c$^{2}$ much heavier than the up and down quarks which have masses in the order of few MeV/c$^{2}$.
There are data on
kaonic atoms available from past experiments (for a review see \cite{Friedman2007}), however the most simple kaonic atoms like kaonic hydrogen and deuterium
are challenging and precision data were obtained in more recent experiments \cite{Iwasaki97,Beer2005,Bazzi2011} using X-ray spectroscopy to unveil the low-energy strong interaction with high precision in these simple systems.

The most basic exotic atom with strangeness represents kaonic hydrogen (K$^{-}$p) which is an electromagnetically bound exotic atom consisting of a proton and an antikaon (K$^{-}$). In this system one can study the explicit and spontaneous chiral symmetry breaking in a fairly direct way by spectroscopy of x-rays emitted in the transitions to the 1s ground state, in this way the threshold data can be deduced.
The K-p interaction at threshold is strongly influenced by the sub-threshold resonance $\Lambda$(1405) which is an interesting hadronic object with a
non-trivial nature \cite{Hyodo2012}. In order to deduce the experimental values of isospin-separated antikaon-nucleon scattering lengths one needs the hadronic shift $\epsilon_{1s}$ and width $\Gamma_{1s}$ in kaonic hydrogen and kaonic deuterium which can be extracted from the X-ray transitions to the 1s ground state in both kaonic systems.

\section{Kaonic hydrogen}
\label{sec-1}
The SIDDHARTA experiment at the DA$\Phi$NE electron-positron collider succeeded to measure the X-ray spectrum of kaonic hydrogen by using an array of silicon drift detectors. From the K transitions the experimental values of $\epsilon_{1s}$ and width $\Gamma_{1s}$ were determined. The energy shift  $\epsilon_{1s}$ is given by the deviation of
the measured K (np $\rightarrow$ 1s) transition energy E$_{np\rightarrow1s}^{meas.}$ from the calculated electromagnetic value E$_{np\rightarrow1s}^{em}$.

\begin{equation}\label{shift}
    \epsilon_{1s} = E_{np\rightarrow1s}^{meas.} - E_{np\rightarrow1s}^{em}
\end{equation}

With a modified Deser formula Equ.\ref{Deser} \cite{meis06}  the antikaon-nucleon scattering length a$_{p}$ can be calculated which is the averaged sum of the a$_{0}$ (isospin I=0) and a$_{1}$ (isospin I=1) scattering lengths (a$_{p}$ = $\frac{1}{2}$(a$_{0}$+a$_{1}$)).

\begin{equation}\label{Deser}
\epsilon_{1s}+\frac{i}{2}\Gamma_{1s} =2\alpha^{3}\mu_{c}^{2} a_{p} (1-2\alpha\mu_{c}(ln\alpha-1)a_{p}),
\end{equation}

It is clear that one has to study the antikaon-neutron interaction by using kaonic deuterium to determine the a$_{0}$ and a$_{1}$.

\section{Kaonic deuterium}
\label{sec-1}
\begin{figure}
\centering
\includegraphics[width=8cm,clip]{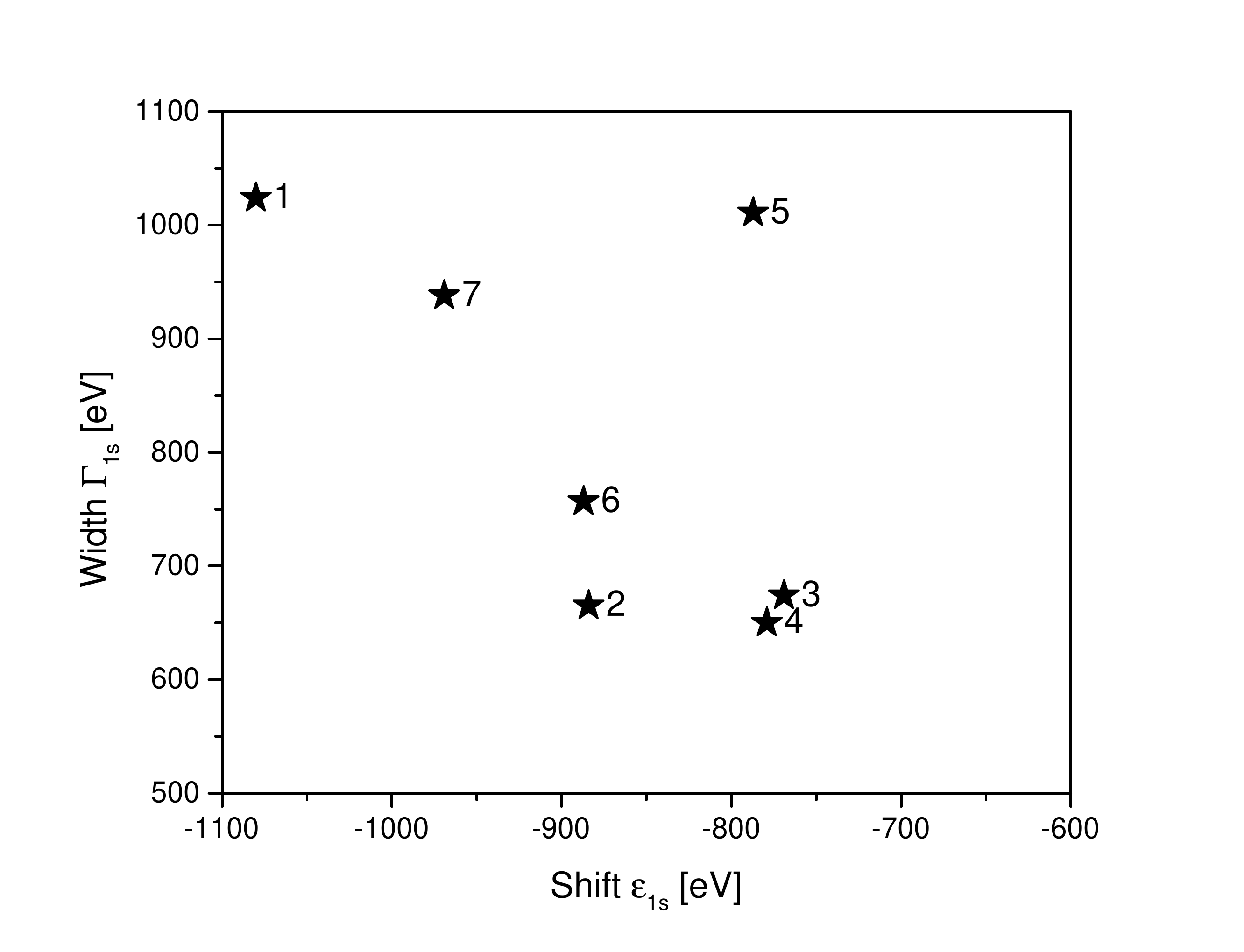}
\caption{Predicted values of $\epsilon_{1s}$ and $\Gamma_{1s}$ obtained in theoretical studies of kaonic deuterium indicated by the numbers 1-7:
1 ref.\cite{oset01}, 2 ref.\cite{meis06}, 3 ref.\cite{gal07}, 4 ref.\cite{meis11}, 5 ref.\cite{shev12}, 6 ref.\cite{miz13}, 7 ref.\cite{weise15}.}
\label{fig-1}       
\end{figure}

The case of kaonic deuterium is more challenging than kaonic hydrogen mainly due to the larger widths of the K lines and the lower X-ray yield expected.
In Fig.\ref{fig-1} kaonic deuterium values of $\epsilon_{1s}$ and $\Gamma_{1s}$ calculated in different theoretical approaches are displayed.

Experimentally the case of kaonic deuterium is still open. SIDDHARTA measured the X-ray spectrum with a pure deuterium filling but due to the limited statistics and the background condition the determination of $\epsilon_{1s}$ and $\Gamma_{1s}$ was impossible. An upper limit for the X-ray yield of the K lines could be extracted from the data: total yield  <0.0143, K$\alpha$ yield < 0.0039 \cite{Cargnelli13}.

\begin{figure}
\centering
\includegraphics[width=8cm,clip]{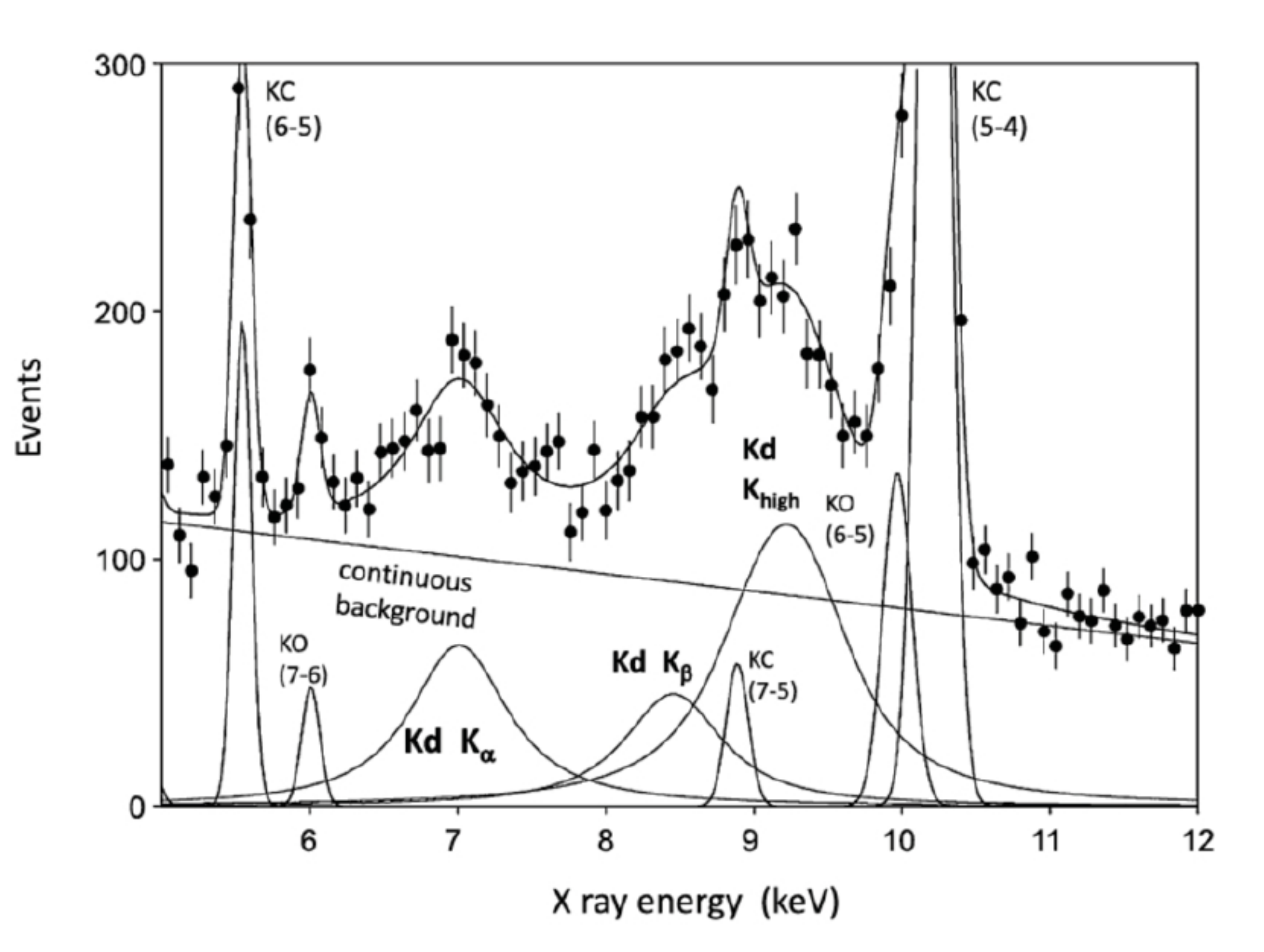}
\caption{Monte-Carlo calculated X-ray spectrum of kaonic deuterium assuming $\epsilon_{1s}$ = -805 eV and $\Gamma_{1s}$ = 750 eV \cite{Cargnelli15}. With this values one gets an estimated
precision of 70 eV in the shift and 150 eV in the width.}
\label{fig-2}       
\end{figure}

A new experiment SIDDHARTA2 is planned which is based on a strongly improved apparatus. The improvements include an
optimized geometry, deuterium gas density, discrimination of K$^{+}$, active shielding and better SDD timing performance by cooling.
According to Monte Carlo studies one expects an X-ray energy spectrum shown in Fig.\ref{fig-2}.

\section{Summary and Outlook}

The SIDDHARTA experiment provided solid data for the antikaon-nucleon interaction at threshold and thus important constraints for the theory of strong interaction with strangeness at low energies. Now a clear description based on an effective field theory with coupled channels is available which is consistent with the information on the antikaon-nucleon interaction \cite{Weise-Ikeda11}. After SIDDHARTA the next important step is the study of the kaonic deuterium X-ray spectrum in order to deduce the isospin separated antikaon-nucleon scattering lengths. This measurement and related studies \cite{Curceanu13} are crucial for the understanding of strangeness and will provide stringent tests of the theoretical description.

\begin{acknowledgement}
We thank C. Capoccia, G. Corradi, B. Dulach, and D. Tagnani from LNF-INFN; and H. Schneider, L.
Stohwasser, and D. Stückler from Stefan-Meyer-Institut, for their fundamental contribution in designing
and building the SIDDHARTA setup. We thank as well the DA$\Phi$NE staff for the excellent working
conditions and permanent support. Part of this work was supported by the European Community-
Research Infrastructure Integrating Activity “Study of Strongly Interacting Matter" (HadronPhysics2,
Grant Agreement No. 227431, and HadronPhysics3 (HP3) Contract No. 283286) under the Seventh
Framework Programme of EU; HadronPhysics I3 FP6 European Community program, Contract No.
RII3-CT-2004- 506078; Austrian Science Fund (FWF) (P24756-N20); Austrian Federal Ministry of
Science and Research BMBWK 650962/0001 VI/2/2009; Croatian Science Foundation under Project No. 1680; Romanian National Authority for Scientific
Research, Contract No. 2-CeX 06-11-11/2006; and the Grant-in-Aid for Specially Promoted Research
(20002003), MEXT, Japan.
\end{acknowledgement}

%

\begin{thebibliography}{}
%
%

\bibitem{Friedman2007} E. Friedman, A. Gal, Phys. Rep.  \textbf{452}, 89 (2007), C.J. Batty, E. Friedman, A. Gal, Phys. Rep. \textbf{287}, 385 (1997).

\bibitem{Iwasaki97} M. Iwasaki et al., Phys. Rev. Lett. \textbf{78}, 3097(1997), T. M. Ito, et al.,
Phys. Rev. C \textbf{58}, 2366 (1998).

\bibitem{Beer2005} G. Beer et al., Phys. Rev. Lett. \textbf{94}, 212302 (2005).

\bibitem{Bazzi2011}M. Bazzi et al., Phys. Lett. B \textbf{704}, 113 (2011), Bazzi, M., Beer, G., Bombelli, L., Bragadireanu, A. M., Cargnelli, M. et al. (SIDDHARTA collaboration), Nucl. Phys. A \textbf{881}, 88 (2012).

\bibitem{Hyodo2012}T. Hyodo and D. Jido, Prog.Part.Nucl.Phys. \textbf{67} (2012) 55-98 arXiv:1104.4474 [nucl-th].

\bibitem{meis06}  U.-G. Meißner, U. Raha, A. Rusetsky, Eur. Phys. J. C \textbf{47}, 473 (2006).

\bibitem{oset01} S.S. Kamalov, E. Oset, A. Ramos, Nucl. Phys. A \textbf{690} 494 (2001).


\bibitem{gal07}  A. Gal, Int. J. Mod. Phys. A \textbf{22}, 226 (2007).

\bibitem{meis11} M. Döring, U.-G. Meißner, Phys. Lett. B \textbf{704}, 663 (2011).

\bibitem{shev12} N.V. Shevchenko, Nucl. Phys. A \textbf{890–891}, 50 (2012).

\bibitem{miz13} T. Mizutani, C. Fayard, B. Saghai, K. Tsushima, Phys. Rev. C \textbf{87}, 035201 (2013).

\bibitem{weise15} W. Weise, Hyp. Int. \textbf{233}, 131 (2015).

\bibitem{Cargnelli13}M. Bazzi [SIDDHARTA], Nucl. Phys. A \textbf{907}, 69 (2013).

\bibitem{Cargnelli15} M. Cargnelli, private communication.


























%
%






















\bibitem{Weise-Ikeda11} Y. Ikeda, T. Hyodo and W. Weise, Phys. Lett. B \textbf{706}, 63 (2011).

\bibitem{Curceanu13} C. Curceanu et al., Nucl. Phys. A \textbf{914}, 251 (2013).

\end{thebibliography}
%
%

\end{document}